\documentclass[twocolumn,showpacs]{revtex4}
\usepackage{amsmath}
\usepackage{amsfonts}
\usepackage{graphicx}

\begin{document}

\title{
Interpretation of Quasielastic Scattering Spectra of Probe Species in Complex Fluids
}

\author{George D. J. Phillies}
\email{phillies@wpi.edu, 508-754-1859}

\affiliation{Department of Physics, Worcester Polytechnic
Institute,Worcester, MA 01609}

\pacs{83.80.Rs,66.10.cg,66.30.hk,83.86.Hf}

\begin{abstract}

The objective of this paper is to correct an error in analyses of quasielastic scattering spectra. The error invokes a valid calculation under conditions in which its primary assumptions are incorrect, leads to misleading interpretations of spectra.  Quasielastic scattering from dilute probes yields the incoherent structure factor $g^{(1s)}(q,t) = \langle \exp(i q \Delta x(t)) \rangle$, with $q$ being the magnitude of the scattering vector ${\bf q}$ and $\Delta x(t)$ being the probe displacement parallel to ${\bf q}$ during a time interval $t$.  The error is a claim that $g^{(1s)}(q,t) \sim \exp(- q^{2} \langle (\Delta x(t))^{2} \rangle/2 )$ for probes in an arbitrary solution, leading to the incorrect belief that $\langle (\Delta x(t))^{2} \rangle$ of probes in complex fluids can be inferred from quasielastic scattering. The actual theoretical result refers only to monodisperse probes in simple Newtonian liquids. In general, $g^{(1s)}(q,t)$ is determined by all even moments $\langle (\Delta x(t))^{2n} \rangle$, $n = 1, 2, 3,\ldots$ of the displacement distribution function $P(\Delta x,t)$.  Correspondingly, $\langle (\Delta x(t))^{2} \rangle$ cannot in general be inferred from $g^{(1s)}(q,t)$.  The theoretical model that ties $g^{(1s)}(q,t)$ to $\langle (\Delta x(t))^{2} \rangle$ \emph{also quantitatively determines} exactly how $\langle (\Delta x(t))^{2} \rangle/2)$ must behave, namely $\langle (\Delta x(t))^{2} \rangle$ must increase linearly with $t$. \emph{If the spectrum is not a single exponential in time}, $g^{(1s)}(q,t)$ does not determine $\langle (\Delta x(t))^{2} \rangle$.

\end{abstract}

\maketitle

\section{Introduction}

Quasielastic scattering of light, x-rays, and neutrons is a powerful experimental tool for the study of complex fluids. For dilute solutions in simple liquids, quasielastic scattering has analytic applications for particle sizing\cite{phillies1990z}.  In non-dilute and more complex systems, quasielastic scattering reveals consequences of intermacromolecular interactions. Four decades ago,  Hallett and co-workers\cite{gray1974a,turner1976a} made a major extension of quasielastic scattering methodology.  To hyaluronic acid and dextran solutions, they added trace concentrations of polystyrene latex spheres. Quasielastic scattering revealed the diffusion of these probe spheres through the polymer solutions. The subsequent four decades have seen enormous extension of probe diffusion methods\cite{phillies2011a}, including studies of probes in highly viscous simple liquids\cite{phillies1981z}, polymer melts\cite{lin1986a}, chemically cross-linked gels\cite{schmidt1989b}, surfactant solutions\cite{phillies1993b}, protein solutions\cite{ullmann1985a}, and the interior of living cells\cite{lubyphelps1987a}. More recently, quasielastic x-ray scattering has extended the range of distance scales over which diffusion can be observed\cite{dierker1995a}.

Probe diffusion has also been studied by other physical techniques, each sensitive to its own characteristic time and distance scales.  For example, fluorescence correlation spectroscopy\cite{magde1972a}, which by varying the probe labelling fraction can measure both the self diffusion coefficient and the mutual diffusion coefficient of the labeled species\cite{phillies1975a,scalettar1989a}, has in recent years been extensively applied to tracer diffusion. Recent work using probe diffusion is sometimes termed \emph{microrheology}, the term microrheology referring to a particular model\cite{mason1996a} for interpreting scattering spectra. In some studies, probe diffusion is viewed as being of interest because it measures viscoelastic properties of the solution.  In other studies, probe diffusion is viewed as being of interest because it measures solution properties that are different from the viscoelastic properties of the solution.

A valuable complement to diffusion studies is provided by measurements on probes being driven by external forces. The overwhelming majority of work on driven probe motion in complex fluids relies on capillary electrophoresis.  The emphasis of this work is on separating charged species. However, separations using, e.g., neutral polymer solutions as support media, equally give information on the dynamics of the support medium\cite{phillies2011a,phillies2012e}. A substantial literature exists on ultracentrifugal sedimentation in complex fluids\cite{laurent1963a,ye1998a}.  A few experiments used magnetic\cite{hough1999a,schmidt2000a} or optical\cite{amblard1996a} tweezers to create oscillatory or other driving forces. Tweezer experiments are particularly interesting because the experimenter can separately control two of the three: drive force, drive frequency, and particle displacement. An alternative complement to probe diffusion is provided by probes in complex fluids in which the fluid itself performs driven motion, e.g., shear\cite{tapadia2006a}.

Quasielastic photon scattering is studied by measuring the intensity-intensity time correlation function
\begin{equation}
      S(q,t) = \langle I(q,\tau) I(q, \tau+t) \rangle.
     \label{eq:Sqtdef}
\end{equation}
Here $q$ is the magnitude of the scattering vector, $I(q,\tau)$ and $I(q, \tau+t)$ are the scattering intensities over short time intervals near $\tau$ and $\tau+t$, and the brackets $\langle \cdots \rangle$ represent an average. In cases of interest here, scattering arises from probe particles surrounded by a complex medium. If the volume being observed is much larger than the volumes over which particle positions and displacements are correlated, quasielastic scattering corresponds as shown by Crosignani, et al.\cite{crosignani1975a} to the intermediate structure factor (or field correlation function) $g^{(1)}(q,t)$ via
\begin{equation}
     S(q,t) = A |g^{(1)}(q,t)|^{2} + B.
     \label{eq:Sqg1def}
\end{equation}
In this equation $A$ and $B$ are constants determined by details of the experimental apparatus; these constants have no effect on the time dependence. Homodyne rather than heterodyne detection of the scattered light is assumed. Crosignani, et al.\cite{crosignani1975a} supply conditions under which the factorization is valid:  The scattering occurs in a volume $V$, within which particle positions are correlated within volumes $\approx V_{s}$ and particle motions are correlated within volumes $\approx V_{d}$.  The factorization proceeds if $V/V_{s}$ and $V/V_{d}$ are both $\gg 1$. The factorization of $S(q,t)$ into $g^{(1)}(q,t)$ is sometimes termed the ``Gaussian approximation''.  This Gaussian approximation is not related to the Gaussian approximation for the particle displacements, as discussed below.

The intermediate structure factor is in turn determined by the time-dependent positions of the scattering particles via
\begin{equation}
    g^{(1)}(q,t) = \left\langle \sum_{i=1}^{N} \sum_{j=1}^{N} \exp(\imath {\bf q} \cdot ({\bf r}_{i}(t+\tau) -  {\bf r}_{j}(\tau) )) \right\rangle
    \label{eq:g1qgeneral}
\end{equation}
In this equation, sums on $i$ and $j$ proceed separately over all $N$ particles in the system, while ${\bf r}_{i}(t+\tau)$ and ${\bf r}_{j}(\tau)$ are the locations of scatterers $i$ and $j$ at times $t+\tau$ and $\tau$, respectively.

In applying eq \ref{eq:g1qgeneral}, two particularly interesting experimental circumstances are described as measuring mutual diffusion or probe diffusion.  Quasielastic scattering on a binary solvent: scatterer system measures the mutual diffusion coefficient, which describes the motion of scatterers down a concentration gradient\cite{phillies1974a,phillies1974b}.  Tracer diffusion experiments examine ternary solvent: matrix : probe systems.  The matrix component is substantially responsible for the system's rheological properties, but is nearly optically inert.  Conversely, the probe component is dilute, has virtually no effect on the rheological properties of the system, but dominates scattering. If matrix scattering is not entirely negligible, there are established, reliable ways to isolate probe scattering, based on spectral subtraction at the level of the field correlation function, as discussed in the Appendix.

Probe particles very nearly do not interact with each other, so the field correlation function for probes reduces (up to normalization constants) to the incoherent scattering function
\begin{equation}
   g^{(1s)}(q,t) =  \langle \exp(\imath q \Delta x(t))\rangle.
 \label{eq:g1sandr}
\end{equation}
with $\Delta x(t)$ being the component parallel to ${\bf q}$ of $\mathbf{\Delta r}(t) = {\bf r}_{i}(t+\tau) -  {\bf r}_{i}(\tau)$.  Probe motions perpendicular to $\mathbf{q}$ do not contribute to $g^{(1s)}(q,t)$. In eq \ref{eq:g1sandr}, the superscript "s" stands for "self".  The term "self" refers to the double sum in eq \ref{eq:g1qgeneral}, in which the $i=j$ terms are the self terms that give rise to eq \ref{eq:g1sandr}.  The $i \neq j$ terms are the "distinct" terms. In moving from eq \ref{eq:g1qgeneral} to eq \ref{eq:g1sandr}, terms of eq \ref{eq:g1qgeneral} in which $i \neq j$ were taken to average to zero, because the relative positions of dilute probes are uncorrelated. An expression formally identical to eq  \ref{eq:g1sandr} describes diffusion measurements using pulsed-field-gradient nuclear magnetic resonance, though with this method $q$ has an entirely different meaning, namely in the simplest case $q = \gamma \delta g$, where $\gamma$ is the gyromagnetic ratio, $\delta$ is a pulse width, and $g = dB/dz$ is the field gradient.

Eqs \ref{eq:g1qgeneral} and \ref{eq:g1sandr} may formally be phrased as averages over displacement distribution functions such as $P(\Delta x, t)$, which gives the time-dependent probability that a scattering particle will displace through $\Delta x$ during time $t$. Two previous papers\cite{phillies2005a,phillies2012a} examined how $g^{(1s)}(q,t)$ and $g^{(1)}(q,t)$ are actually related to the displacement distribution functions. Their extended derivations are not repeated here.  The two prior papers were primarily concerned with establishing formal relationships between dynamic structure factors and probabilities for scatterer displacements. The significance of these relationships for interpretating experiments was at most a secondary consideration.  This paper focuses on interpreting experiments.

Section II of this paper presents the correct general relationship between $g^{(1s)}(q,t)$ and $P(\Delta x, t)$.  Section III discusses the special case of probe particles in a purely Newtonian fluid.  Section IV notes experimental findings bearing on the relative significance of Sections II and III.  Section V considers paths for interpreting probe diffusion spectra. Section VI treats the determination of $P(\Delta x, t)$, relationships between  $g^{(1s)}(q,t)$ and trapping/hopping behavior, and, to close on a positive note, several cases in which quasielastic scattering from diffusing probes, correctly interpreted, has given valuable information about complex fluids and the objects diffusing in them. An Appendix discusses spectral subtraction at the field correlation level.

\section{General Case\label{sectiongeneralcase}}

This section summarizes what $g^{(1s)}(q,t)$ reveals about particle displacements. The self part of the intermediate structure factor is determined by the normalized displacement distribution function $P(\Delta x, t)$, namely the average in eq \ref{eq:g1sandr} is
\begin{equation}
       g^{(1s)}(q,t) = \int_{-\infty}^{\infty}  d(\Delta x)  \exp(i q \Delta x) P(\Delta x, t).
        \label{eq:g1sPDelta}
\end{equation}

On taking a Taylor series expansion of the exponential in powers of $q$, reflection symmetry, namely $P(\Delta x, t) = P(-\Delta x, t)$, eliminates all terms odd in $q$.  As a result, $g^{(1s)}(q,t)$ and its logarithm are necessarily power series in $q^{2}$.  The coefficients of the $q^{2n}$ are generated by the even moments $\langle(\Delta x)^{2n}\rangle$ of $P(\Delta x, t)$.  The lead terms of the expansion for $g^{(1s)}(q,t)$ are\cite{phillies2005a}
\begin{displaymath}
      g^{(1s)}(q,t) = N \exp\left( - \frac{1}{2} q^2 \langle ( \Delta x(t))^{2} \rangle + \frac{1}{24} q^{4}( \langle (\Delta x(t))^{4} \rangle\right.
\end{displaymath}
\begin{equation}
            \left. - 3\langle (\Delta x(t))^{2} \rangle^{2}) - {\cal O}(q^{6})\right).
      \label{eq:g1sanddisplacements}
\end{equation}
The complete expansion requires all even moments $\langle(\Delta x)^{2n}\rangle$.

It was early shown that quasielastic scattering from a binary solvent: macromolecule system determines via $g^{(1)}(q,t)$ the mutual diffusion coefficient\cite{phillies1974a,phillies1974b}.  Theoretical approaches to computing $g^{(1)}(q,t)$ and the mutual diffusion coefficient of non-dilute colloid solutions have historically followed routes very different from routes based on the displacement distribution function. Only very recently\cite{phillies2012a} was a solution for $g^{(1)}(q,t)$ in terms of displacement distribution functions obtained. In this solution, the expansion of eq \ref{eq:g1qgeneral} was shown to require averages over two different displacement distribution functions, namely $P(\Delta x, t)$ and a new distribution function $P_{2}(\Delta x, t, \mathbf{R}_{12})$. $P_{2}$ is a two-particle conditional displacement distribution function, in which $\Delta x$ is the displacement of particle $1$ during $(0, t)$ given that the vector $\mathbf{R}_{12}$ from particle $1$ to some particle $2$ at time $0$ has a given value.

\section{Special Case: Probes in Simple Newtonian Liquids \label{sectionsimple} }

The earliest quasielastic scattering experiments were performed on dilute suspensions of monodisperse scattering particles in simple Newtonian solvents. Cummins, et al.'s results on polystyrene spheres in water\cite{cummins1964a} are the archetype.  The resulting spectra were interpreted by invoking a mechanical model for the motions of diffusing particles. The mechanical model was provided by the Langevin equation, which in one dimension is
\begin{equation}
      m\frac{d^{2} x(t) }{dt^{2}} = - f \frac{dx(t)}{dt} +{\cal F}_{x}(t).
     \label{eq:Langevin}
\end{equation}
Here $x(t)$ is a coordinate of the diffusing particle, $m$ is the particle mass,  $f$ is the particle's drag coefficient, and ${\cal F}_{x}(t)$ is the random force, called \emph{random} because in the Langevin model the values of ${\cal F}_{x}(t)$ at different instants in time are uncorrelated. Within the model, ${\cal F}_{x}$ cannot be predicted beyond stating that ${\cal F}_{x}$ has certain statistical properties.

The canonical literature treatment of the Langevin model as applied to quasielastic light scattering is the volume by Berne and Pecora\cite{berne1976a}, notably their Section 5.9. Berne and Pecora show that the Langevin model is appropriate for polystyrene spheres in water, on the time and distances scales observed by quasielastic light scattering. From the Langevin model and the requirement that the system remains in thermal equilibrium, a series of conclusions about the statistical properties of the particle motion follow. In particular:
\begin{description}

\item[(i)] The mean-square average value of ${\cal F}_{x}(t)$ must be consistent -- the fluctuation-dissipation theorem -- with the drag coefficient $f$ and the thermal energy $k_{B}T$.

\item[(ii)] The distribution of particle displacements during a time interval $\Delta t$ is the same for all time intervals $(t, t+\Delta t)$.

\item[(iii)] Velocity correlations are evanescent. For time steps appreciably longer than $m/f$, which for Brownian particles is actually a quite short time, particle displacements in a series of time steps are very nearly independent from each other.

\end{description}

Conclusion (ii) corresponds to the statement that $x(t)$ is the sum of a series of identically-distributed random variables. Conclusion (iii) corresponds to the independent statement that the time evolution of $x(t)$ is described by a Markoff process.  In this very special case, the distribution of particle displacements is described by Doob's Theorem\cite{doob1942a}, which is closely related to the central limit theorem.  Doob's theorem treats random processes such as $\Delta x(t)$, while the central limit theorem treats random variables.  For the Langevin model, Doob's Theorem shows that the distribution of particle displacements is a Gaussian
\begin{equation}
    P(\Delta x) =   \left(2\pi  \langle (\Delta x)^{2} \rangle \right)^{-1/2} \exp(- (\Delta x(t))^{2}/ 2 \langle (\Delta x)^{2} \rangle ),
    \label{eq:gaussianform}
\end{equation}
and, separately, that successive displacements have a joint Gaussian distribution.  For this special case, the incoherent scattering function reduces to
\begin{equation}
     g^{(1s)}(q,t) = \exp(- q^{2} \langle (\Delta x(t))^{2} \rangle/2).
 \label{eq:g1swrong}
\end{equation}
Equation \ref{eq:g1swrong} is accurate for the systems considered by Berne and Pecora\cite{berne1976a}, namely highly dilute solutions of monodisperse objects in simple Newtonian solvents.

However, Berne and Pecora\cite{berne1976a}, especially their Appendix 5.A and Section 5.9 leading to their eq 5.9.6, also prove the other important consequence of the Langevin model and Doob's theorem: The Langevin model determines the exact value of $\langle (\Delta x(t))^{2} \rangle$.  On time and distance scales accessible to quasielastic scattering, the Langevin model requires
\begin{equation}
     \langle (\Delta x(t))^{2} \rangle = 2 D t.
     \label{eq:meansquare}
\end{equation}
$D = k_{B}T/f$ is the diffusion constant, a quantity \emph{that does not depend on time}. Time independence of $D$ is forced by the calculation, because $D$ results from a time integral over ($0 \leq t \leq \infty$). Here $k_{B}$ is Boltzmann's constant and $T$ is the absolute temperature.

Equations \ref{eq:g1swrong} and \ref{eq:meansquare} come as a package; they are equally consequences of the Langevin model. Correspondingly, Berne and Pecora show for diffusing monodisperse Brownian particles that the Langevin model requires that the field correlation function is a simple exponential
\begin{equation}
      g^{(1s)}(q,t) = \exp(- q^{2} D t).
      \label{eq:g1ssimple}
\end{equation}
For unclear reasons -- the literature error noted in the introduction -- Berne and Pecora's entirely correct Chapter 5 is being misread as proving that eq \ref{eq:g1swrong} is always correct, even when the time relaxation of $g^{(1s)}(q,t)$ is not a simple exponential. Berne and Pecora in fact prove exactly the opposite. The valid contrapositive of their valid result is: If the relaxation is not a single exponential, then the Langevin model must not be applicable to the system, and therefore invocation of the Langevin Model prediction eq \ref{eq:g1swrong} is invalid.

Berne and Pecora's discussion refers only to particle motions described by the Langevin equation, for which eqs \ref{eq:gaussianform}-\ref{eq:g1ssimple} are all correct. Their discussion corresponds to many experiments that were of interest when they were writing, notably particle sizing studies\cite{dahneke1983}. For dilute monodisperse colloids, the $t$ and $q$ dependences of $g^{(1s)}(q,t)$ are precisely as predicted by the Langevin model, in particular $g^{(1s)}(q,t) \sim  \exp(- \Gamma t)$ with $\Gamma \propto q^{2}$.

If eq \ref{eq:gaussianform} and the Langevin equation described the particle motions, then the spectrum would necessarily be a simple exponential in $q^{2}t$.  If the decay of the field correlation function is not a simple exponential, then eq \ref{eq:gaussianform} and the Langevin model do not describe how the scattering particles move.  In systems in which the spectrum is more complex than a simple exponential, eq \ref{eq:g1swrong} is invalid.  $\log(g^{(1s)}(q,t))$ only reveals the mean-square displacement of the particles if $g^{(1s)}(q,t)$ is a simple exponential in $t$.

Why does eq \ref{eq:g1sanddisplacements} ever reduce to eq \ref{eq:g1swrong}?  If $P(\Delta x, t)$ is a Gaussian in $\Delta x$, $P(\Delta x, t)$ is entirely characterized by its second moment $\langle (\Delta x)^{2} \rangle$.  For a Gaussian displacement distribution function, the higher moments of $P(\Delta x, t)$ have values such that the coefficients of the higher-order terms ($q^{2n}$ for $n \geq 2$)  of eq \ref{eq:g1sanddisplacements} all vanish. For a Gaussian $P(\Delta x, t)$, the only non-zero part of eq \ref{eq:g1sanddisplacements} is eq \ref{eq:g1swrong}.  This disappearance of the higher-order terms is unique to a Gaussian $P(\Delta x, t)$. For any other $P(\Delta x, t)$, the higher-order terms of eq \ref{eq:g1sanddisplacements} do not vanish.

\section{Experimental Findings \label{sectionexperiment}}

What do experiments say about $P(\Delta x, t)$ and $g^{(1s)}(q,t)$?  There are systems in which the Langevin model is adequate, namely dilute monodisperse particles suspended in simple Newtonian fluids. For probe diffusion in complex fluids, experiment provides a far more complex picture. Consider a few representative experiments:

On relatively long time scales, $P(\Delta x,t)$ is accessible via particle tracking, e.g., experiments by Apgar, et al.\cite{apgar2000a}, Tseng and Wirtz\cite{tseng2001a}, and Xu, et al.\cite{xu2002a} on probe diffusion in glycerol, actin solutions and gels, and gliadin solutions. These authors used video recording and computer image analysis to follow large numbers of particles simultaneously. They report $P(\Delta x, t)$ and $\langle (\Delta x(t))^{2} \rangle$ in their systems.  Probes in glycerol follow eqs \ref{eq:gaussianform} and \ref{eq:meansquare}.  For probes in complex fluids, $P(\Delta x, t)$ has decidedly non-Gaussian forms.  Correspondingly, the mean-square displacement does not increase linearly in time. Experiment thus shows that eq \ref{eq:gaussianform} and \ref{eq:meansquare} are not uniformly correct for probes in polymer solutions.

Quasielastic scattering spectra of probes in polymer solutions are often markedly non-exponential.  For polystyrene latex sphere probes in hydroxypropylcellulose: water, this author and Lacroix\cite{lacroix1997a} found stretched exponentials in time
\begin{equation}
     g^{(1s)}(q,t) =  a \exp(- \theta t^{\beta}) \equiv a \exp(- (t/\tau)^{\beta}).
     \label{eq:gisqeext}
\end{equation}
Here $\beta$ is a scaling exponent while $\theta$ and $\tau$ are prefactors. A series of papers by Streletzky and collaborators on the same chemical system (most recently, ref.\ \onlinecite{phillies2003a}) established by viewing a much wider range of delay times that $g^{(1s)}(q,t)$ is in general a sum of two stretched exponentials in time. For these systems, eqs \ref{eq:gaussianform}-\ref{eq:g1ssimple} are inapplicable.

Finally, I note a very simple model system in which eqs \ref{eq:gaussianform} and \ref{eq:g1swrong} fail. The system is a dilute aqueous dispersion of polystyrene spheres, in which the spheres are of two different sizes.  There are no sphere-sphere interactions.  Each sphere individually performs Brownian motion as described by the Langevin equation. Therefore, for each sphere in the mixture, $P(\Delta x, t)$ is a Gaussian in $\Delta x$, and $\langle (\Delta x)^{2} \rangle$ increases linearly in time.  For the mixture as a whole, the weighted average over all particles of the mean-square displacement must also increase linearly with time. The appropriate weights are the intensities $A_{1}$ and $A_{2}$  of the light scattered by the two species, so that
\begin{equation}
     \frac{d  \langle (\Delta x(t))^{2} \rangle}{ dt} = 2 \frac{A_{1} D_{1} +A_{2} D_{2} }{A_{1}+A_{2}}
     \label{eq:deltaxaverage1}
\end{equation}
Here $D_{i}$ is the diffusion coefficient of species $i$. For Rayleigh-Gans-Debye scattering, up to a constant the $A_{i}$ are
\begin{equation}
       A_{i} = n_{i} M_{i}^{2} P_{i}(q) =  c_{i} M_{i}P_{i}(q).
      \label{eq:scatteringintensity}
\end{equation}
Here $n_{i}$ is the number density of scatterer $i$, $M_{i}$ is the molecular weight of scatterer $i$, $c_{i}$ is the mass concentration of species $i$, and $P_{i}(q)$ is the static structure factor of species $i$\cite{tanford1961a}.

If eq \ref{eq:g1swrong} were correct, one would have
\begin{equation}
     \frac{d  \langle (\Delta x(t))^{2} \rangle}{ dt} = -\frac{2}{q^{2}} \frac{d \log(g^{(1s)}(q,t)}{dt}
    \label{eq:useeqg1swrong}
\end{equation}
We now calculate explicitly the RHS of this equation. To simplify the discussion, the larger spheres are identified as species 2. The mixture's field correlation function is
\begin{equation}
      g^{(1s)}(q,t) = A_{1} \exp(-D_{1} q^{2} t) + A_{2} \exp(-D_{2} q^{2} t).
      \label{eq:doubleexponential}
\end{equation}
The logarithmic derivative of eq \ref{eq:doubleexponential} is
\begin{displaymath}
     - \frac{2}{q^{2}} \frac{ d \ln( g^{(1s)}(q,t))} {dt} =
  \end{displaymath}
\begin{equation}
        2 \frac{A_{1} D_{1} \exp(-D_{1} q^{2} t) + A_{2} D_{2} \exp(-D_{2} q^{2} t)}{A_{1} \exp(-D_{1} q^{2} t) + A_{2} \exp(-D_{2} q^{2} t)}
      \label{eq:meansquare2}
\end{equation}
At short times, $\exp(- D q^{2} t) \approx 1$ for both species and
\begin{equation}
     - \frac{2}{q^{2}} \frac{ d \ln( g^{(1s)}(q,t))} {dt} = \frac{A_{1} D_{1} + A_{2} D_{2}}{A_{1}  + A_{2} }.
      \label{eq:meansquare3}
\end{equation}
The initial slope of eq \ref{eq:meansquare2} thus agrees with the initial slope of eq \ref{eq:deltaxaverage1}.

The long-time behavior of eq \ref{eq:meansquare2} is quite different, as made clear by factoring an $\exp(- D_{2} q^{2} t)$ from its numerator and denominator, so
\begin{equation}
- \frac{2}{q^{2}} \frac{ d \ln( g^{(1s)}(q,t))} {dt} = 2 \frac{A_{1} D_{1} \exp(-(D_{1}-D_{2}) q^{2} t) + A_{2} D_{2}}{A_{1} \exp(-(D_{1}-D_{2}) q^{2} t) + A_{2}}
      \label{eq:meansquare4}
\end{equation}
$D_{2} < D_{1}$; the exponentials decay at large time. With increasing $t$, the fast relaxation ceases to contribute to $g^{(1s)}(q,t)$, so that
\begin{equation}
    \lim_{t \rightarrow \infty} - \frac{2}{q^{2}} \frac{ d \ln( g^{(1s)}(q,t))} {dt} = 2 D_{2}
          \label{eq:meansquare5}
\end{equation}
At large times, only the larger spheres contribute to $d \ln( g^{(1s)}(q,t))/dt$. We have now calculated the rate of increase in the mean-square displacement as inferred from eq \ref{eq:g1swrong}.  The general result is eq \ref{eq:meansquare2}.  At short times, eq \ref{eq:g1swrong} leads to eq \ref{eq:meansquare3}, which is the correct result of eq \ref{eq:deltaxaverage1}. The short-time result is the same as Koppel's demonstration for dilute particles that the first time cumulant of the spectrum gives the average diffusion coefficient\cite{koppel1972a}.  However, at large times eq \ref{eq:g1swrong} leads to eq \ref{eq:meansquare4}.  At large times, only the slower-moving species contributes to the $d  \langle (\Delta x(t))^{2} \rangle/dt$ predicted by eq \ref{eq:g1swrong}.  An experimenter who used eq \ref{eq:g1swrong} to interpret spectra on this binary sphere mixture would conclude that the diffusing particles moved more rapidly at early times, but slowed down at later times, the so-called "sub-diffusion" phenomenon.  Here sub-diffusion is an artifact.  At long times, the smaller spheres continue to move, but they do not contribute to $g^{(1s)}(q,t)$. At long times, the contribution of the more rapidly-moving species to $g^{(1s)}$ is negligible, so the nominal mean-square displacement inferred from eq \ref{eq:g1swrong} is entirely determined by scattering from the slower spheres.

Experiment thus demonstrates that neither eq  \ref{eq:gaussianform} nor eq \ref{eq:g1ssimple} is generally valid for probe diffusion in complex fluids. Even in a Newtonian fluid, a model system in which $g^{(1s)}(q,t)$ does not decay exponentially in time does not follow eqs \ref{eq:g1swrong} and \ref{eq:meansquare2}, exactly as required by Doob's theorem. Interpretations of quasielastic scattering spectra for probes in complex fluids, based on the Gaussian approximation of eq \ref{eq:g1swrong}, are therefore incorrect. Interpretations of spectra of probes in complex fluids, in terms of particle displacements, are properly based on eq \ref{eq:g1sanddisplacements}, which correctly reflects the non-Gaussian displacement distribution function of real probes in these systems.

\section{Interpretations of Quasielastic Scattering Spectra \label{sectioninterpretations}}

First, every physical $g^{(1s)}$, viewed only as a function of $t$, corresponds to a system in which the mean-square particle displacement increases linearly with time. However, the correspondence is not unique. The same $g^{(1s)}(q,t)$ may also correspond to systems in which particle thermal motions are more complex.  In consequence, from a $g^{(1s)}(q,t)$ measured over a full range of times and a single $q$, one cannot infer how the particle displacement depends on time.

This result has a purely mathematical basis, namely that the field correlation function can always be represented via a Laplace transform as
\begin{equation}
     g^{(1s)}(q,t) = \int_{0}^{\infty} A(\Gamma) \exp(- \Gamma t).
    \label{eq:laplace}
\end{equation}
Here $\Gamma$ is a relaxation rate and $A(\Gamma)$ is the contribution of relaxations having decay rate $\Gamma$ to $g^{(1s)}(q,t)$.  So long as the system does not have relaxational modes with negative amplitudes, $A(\Gamma)$ is everywhere positive or zero. In this case, there is always a system having the same $g^{(1s)}(q,t)$ as the system of interest, and in which $ \langle (\Delta x(t) )^{2} \rangle$ increases linearly in time.  The system can be physically constructed as a mixture of polystyrene spheres having all different sizes.  The composition of the mixture is determined by $A(\Gamma)$: One adds to the mixture just enough polystyrene spheres having diffusion coefficient $\Gamma/q^{2}$ so that their contribution to the scattering spectrum is $A(\Gamma)$.  For each sphere, $\langle (\Delta x(t) )^{2} \rangle$ increases linearly in time, so therefore $\langle (\Delta x(t) )^{2} \rangle$ of the mixture also increases linearly in time.  Thus, an arbitrary (except for the weak requirement $A(\Gamma) \geq 0 \ \forall \ \Gamma$) form for $A(\Gamma)$ corresponds as one non-unique possibility to a system in which $\langle (\Delta x(t) )^{2} \rangle$ increases linearly in time.

It has repeatedly been found that $g^{(1s)}(q,t)$ decays in time as the stretched exponential of eq \ref{eq:gisqeext}. If one interpreted this time dependence by applying eq \ref{eq:g1swrong}, one would conclude
\begin{equation}
        \frac{q^{2}}{2} \langle (\Delta x(t) )^{2} \rangle = \theta t^{\beta}.
         \label{eq:subdiffusive}
\end{equation}
In the common case $\beta < 1$, from eq \ref{eq:subdiffusive} one would infer that the mean-square particle displacement increases less rapidly at large times than at small times. The inference is incorrect.  A more reasonable interpretation for $\beta <1$ is that diffusion in the complex fluid is created by modes having a range of relaxation times, some longer than others, the contribution of the slower modes to the spectrum becoming more important at longer times.

It is not suggested that there does not exist subdiffusive motion.  Such motion has unambiguously been observed experimentally. Amblard, et al.,\cite{amblard1996a} studied probe motion in f-actin solutions using video microscopy.  Small-bead motion was diffusive; larger-bead diffusion was subdiffusive with $\beta \approx 3/4$.  However, Amblard, et al.'s work was not based on quasielastic scattering. Whenever particle motion is subdiffusive, the light scattering spectrum will not be a simple exponential. The scattering spectrum will follow eq \ref{eq:g1sanddisplacements}, showing that the relationship between  $g^{(1s)}(q,t)$ and $\langle (\Delta x(t) )^{2} \rangle$ is a neither-way street.  Just as one cannot in general calculate $\langle (\Delta x(t) )^{2} \rangle$ from $g^{(1s)}(q,t)$, so also one cannot in general calculate $g^{(1s)}(q,t)$ from $\langle (\Delta x(t) )^{2} \rangle$. To calculate $g^{(1s)}(q,t)$, all higher moments of $P(\Delta x)$ are needed.

\section{Discussion\label{sectiondiscussion}}

The primary objective of this paper was to correct a belief that that $g^{(1s)}(q,t)$ can in general be used to determine the mean-square displacement of probes in complex fluids.  The belief appears to have arisen from a misreading of Chapter 5 of Berne and Pecora's excellent monograph, in which Berne and Pecora discuss the motions of monodisperse probe particles in simple Newtonian fluids under the Langevin model. The spectra of monodisperse Langevin-model particles are necessarily single exponentials. The calculation in Berne and Pecora is correct, but does not refer to non-Newtonian fluids or polydisperse scatterers.

The actual functional form of $P(\Delta x, t)$ can be inferred, at least approximately, from the angular dependence of $g^{(1s)}(q,t)$.  Eq \ref{eq:g1sPDelta} shows that the correlation function $g^{(1s)}(q,t)$ is the spatial Fourier transform of $P(\Delta x, t)$.  If $g^{(1s)}(q,t)$ is determined sufficiently accurately over an adequate range of $q,$ an inverse spatial Fourier transform takes the experimenter back from $g^{(1s)}(q,t)$ to $P(\Delta x, t)$.  To the author's knowledge, this inversion has only been done for the ubiquitous polystyrene spheres in distilled water, for which $g^{(1s)}$ is a simple exponential $\exp(- \Gamma t)$ with $\Gamma$ accurately linear in $q^{2}$. $P(\Delta x, t)$ in this system has the expected Gaussian form.  Measurements of the $q$-dependence of $g^{(1s)}$ for probes in complex fluids are less common, though note, e.~g., Streletzky and Phillies\cite{streletzky1998q} on probes in hydroxypropylcellulose: water. Their solutions may be viewed as highly complex fluids with issues arising from among other things polydispersity, variations in substitution, a liquid-liquid crystal phase transition at larger concentration, and a pseudotheta transition with increasing temperature, not to mention that these solutions are viscous and viscoelastic. The probe spectra reflect the high complexity. These authors found spectra having multiple relaxational modes, some of which had relaxation rates that did not scale linearly in $q^{2}$, proving that these modes did not correspond to Gaussian displacement distribution functions.

Particle tracking is sometimes used to generate the simplified $\langle (\Delta x(t))^{2} \rangle$, rather than the full $P(\Delta x, t)$. The simplification is potentially hazardous. If one has not determined $P(\Delta x, t)$ one does not know if the particle motion process corresponds to simple diffusion.  Any physically reasonable $P(\Delta x, t)$ has some second moment $\langle (\Delta x(t))^{2} \rangle$, but if the form of $P(\Delta x, t)$ is unknown, one cannot tell if it is meaningful to characterize $P(\Delta x, t)$ by its second moment or by the corresponding nominal diffusion coefficient
\begin{equation}
      D(t) = \langle (\Delta x(t))^{2} \rangle/ 2 t.
      \label{eq:timedependentD}
\end{equation}
In a complex fluid, characterization of probe motions via measurement of the second moment $\langle (\Delta x(t))^{2} \rangle$ may well be inadequate.

The assertion that the central limit theorem guarantees that $P(\Delta x, t)$ is a Gaussian in $\Delta x$ is sometimes describes as the "Gaussian Approximation".  Experiments such as those summarized above prove that this assertion is incorrect.  The central limit theorem (for random variables) and Doob's theorem (for random processes) are well known.  Where do their invocations go wrong? The central limit theorem and Doob's theorem are statements about the sum of a large number of identically distributed, independent random processes. As applied to the diffusion problem, the displacement $\Delta x$ during experimentally accessible times can be expressed as the sum of a large number of far smaller steps $\delta x$, each taken during a far smaller time interval $\delta t$. If a random variable $\Delta x$ is the sum of a large number of identically distributed \emph{independent} variables, i.e., if $\delta x(t)$ is a Markoff process, it is in general the case that $P(\Delta x, t)$ is a Gaussian in $\Delta x$. This rationale fails because it refers to a sum of \emph{independent} random processes.  The process that generates $\delta x$ for probes in a viscoelastic fluid is not a Markoff process, because the system has memory.  The central limit theorem and Doob's theorem are therefore not applicable. For probes in complex fluids, the processes generating the steps $\delta x$ are highly correlated, because the "random" forces that determine the $\delta x(t)$ are controlled by the shear modulus $G(t)$, which in a complex fluid has a long correlation time.  Correspondingly, the time correlation function of the random force $\langle {\cal F}_{x}(0) {\cal F}_{x}(t) \rangle$ is long-lived, not $\sim \delta(t)$.  The friction force $f \dot{x}(t)$ of the Langevin equation is replaced with the memory function $\sim \int ds \langle {\cal F}_{x}(0) {\cal F}_{x}(s) \rangle \dot{x}(t-s)$ of a Mori-Zwanzig equation, as discussed in Berne and Pecora. Even in a Newtonian fluid, the rationale leading to the Gaussian approximation fails for polydisperse Brownian particles, because each particle has memory. It remembers how big it is.

An alternative to the central limit theorem is the \emph{small-$q$ approximation}. The nominal idea in the small-q approximation is that the rhs of eq \ref{eq:g1sanddisplacements} is a power series in $q$. If one went to sufficiently small $q$, one might hope that the $q^{2}$ term in the exponential would become dominant, so that eq \ref{eq:g1swrong} would approach being valid.  This hope is not met.  For the simplest case of a mixture of diffusing particles, $g^{(1s)}(q,t)$ is in fact a power series in $q^{2} t$.  If one goes to smaller $q$, in order to determine spectra equally accurately one needs to observe the same fractional degree of decay of $g^{(1s)}(q,t)$. One must therefore go out to longer times.  At those longer times, the \cal{O}($q^{4}$) terms are as significant as they were at larger $q$ and smaller $t$ but the same $q^{2} t$.  Said differently, the coefficients of the correct Taylor series (in $q$) expansion of $g^{(1s)}(q,t)$ are time-dependent. In order for the lead term of the expansion to be dominant, the expansion must be limited not only to small $q$ but also to to small $t$.  If $t$ is large, no matter how small $q$ has been made, the higher-order in $q$ terms are as important as the lower-order terms.  Only at small $q$ and small $t$ is a single-term small-$q$ expansion valid.  The valid small-$q$ expansion is $1-q^{2} \langle (\Delta x(t))^{2} \rangle$, which only describes the leading slope of $g^{(1s)}(q,t)$ at small times.

Consider spectra described by eq \ref{eq:gisqeext}.  The exponential in eq \ref{eq:g1sanddisplacements} scales as $q^{2}$ or is a power series in $q^{2}$, so therefore $\theta$ should also be a power series in $q^{2}$, perhaps simply by being linear in $q^{2}$.  Indeed, for probes in in some \emph{but not other} aqueous hydroxypropylcellulose solutions, Streletzky\cite{streletzky1998q} confirmed experimentally $\theta \sim q^{2}$ over a wide range of $q$.  If $\theta$ were replaced with $\tau^{-\beta}$, one would have
\begin{equation}
        \tau \sim q^{-2/\beta}.
        \label{eq:tauq}
\end{equation}
$\beta$ is often in the range 0.5-1.0, so $\tau$ often depends on $q$ as $q^{-3\pm 1}$. If one interpreted $\tau$ to be a relaxation time, the $q$-dependence from eq \ref{eq:tauq} would be strange indeed: The relaxation would occur more rapidly over large distances (small $q$) than over short distances (large $q$).  This strange $q$-dependence is simply an artifact of way one has parameterized $g^{(1s)}(q,t)$, and the identification of $\tau$ as a relaxation time.  In terms of eq \ref{eq:gisqeext}, $(\theta, \beta)$ provides a natural parameterization while $(\tau,\beta)$ is less transparent.  If mean relaxation times are inferred from the spectral time moments
\begin{equation}
      \langle T_{n} \rangle = \int_{0}^{\infty} t^{n} g^{(1s)}(q,t) dt
      \label{eq:Tmoments}
\end{equation}
of $g^{(1s)}(q,t)$, the choice of parameterizations in eq \ref{eq:gisqeext} has no consequences.  The two paramaterizations of $g^{(1s)}(q,t)$ lead to the same $\langle T_{n} \rangle$.

Spectra of diffusing probes showing two relaxations on very different time scales are sometimes interpreted in terms of caging and hopping relaxations.  The notion is that the medium supplies regions of low potential energy within which probes are free to move ("caging").  The regions are separated by barriers of high potential energy, across which probes only pass on rare occasion ("hopping").  The short time-scale relaxation is said to correspond to caging, while the long time-scale relaxation is said to correspond to hopping.

Computer simulation studies by Luo and Phillies and Luo, et al., test the caging-hopping interpretation\cite{luo1995a,luo1996a}.  These simulations represented Brownian particles moving through a square lattice or a random glass of Lennard-Jones force centers.  The force centers were immobile. Probe motions were generated via the Metropolis algorithm.  These studies differed from some earlier work in that they determined not only time dependent mean-square displacements and effective diffusion coefficients but also obtained $P(\Delta r,t)$ and $g^{(1s)}(q,t)$. By varying the nominal temperature, trapping, hopping, and hindered diffusion behaviors were obtained.  At low temperatures, probe particles explored the volume of their traps; after a certain relaxation time $\langle r^{2}(t) \rangle$ ceased to increase.  At high temperatures, $P(\Delta r, t)$ was nearly Gaussian, with $\langle r^{2}(t) \rangle$ increasing linearly in time even at short times.

Luo, et al., evaluated $g^{(1s)}(q,t)$ for $q^{-1}$ extending from a small fraction of the size of a single potential energy minimum out to distances substantially larger than a typical distance between force centers. At low and high temperatures, $g^{(1s)}(q,t)$ showed nearly exponential relaxations, though at small $T$ and small $q$ the relaxation fell to a non-zero baseline. The baseline was non-zero because the particles were permanently trapped in small isolated volumes of the system.  At intermediate temperatures, relaxations were single-exponential at large $q$ but double-exponential at small $q$.  At the same intermediate temperatures, $P(\Delta r, t)$  was radically non-Gaussian, with local maxima and minima created by local potential energy minima, potential energy saddle points, and times required to traverse local energy maxima.

Other, physically different, systems also give bimodal spectra.  In contrast to Luo, et al.'s probes moving though a fixed matrix, in which relaxations are only bimodal for some values of $q$, relaxations of dilute bidisperse suspensions are double-exponential at all $q$. An alternative model system in which monodisperse particles show several very different classes of relaxation behavior is shown by Glotzer, et al.'s\cite{glotzer2000a} computer simulations of three-dimensional glasses, in which one finds distinct long-lived populations of slow and fast-moving particles, with the immobile particles in clumps and the rapidly moving particles lying in thin ribbons.

Thus, in order to distinguish between systems containing species with two different dynamic behaviors, and systems in which there is local trapping with escapes from the traps at longer times, it is experimentally necessary to study $g^{(1s)}(q,t)$ over a wide range of $q$.  Observations at fixed $q$ of double-exponential relaxations do not reveal whether one is seeing trapping with hopping, or whether the system is in some sense dynamically bidisperse.  Furthermore, in the cases in which $g^{(1s)}(q,t)$ was observed by Luo, et al., to be very nearly the sum of two exponentials, $P(\Delta r,t)$ on interesting distance scales had an elaborate dependence on $r$ with multiple maxima and deep minima.  The interpretation that a biexponential $g^{(1s)}(q,t)$ must correspond to a $P(\Delta r,t)$ that is a sum of two Gaussians, each with a mean-square width increasing linearly in time, is disproved by Luo, et al.'s results. These issues here have been discussed with respect to quasielastic scattering, but as noted above pulsed-field-gradient NMR measures the same mathematical aspect of particle displacements, and may therefore have similar issues appropriate to its time and distance scales.

Finally, the observation that quasielastic scattering does not determine the mean-square probe displacement certainly does not mean that probe diffusion is ineffective.  Probe diffusion measurements can certainly be used to obtain novel information about complex fluids. The richness of the revealed information corresponds to the depth with which models for probe motion are constructed. As a positive conclusion, two successful applications of probe diffusion are noted:

(i) A long-time question in the study of surfactant solutions is the determination of the aggregation number $n$ of surfactant molecules in micelles.  One of many approaches to this question has been to use quasielastic scattering to determine an effective hydrodynamic radius of the micelles.  Perhaps after some hydrodynamic modeling to account for micelle shape, spherical micelles being the simplest case, the measured diffusion coefficient can be transformed to an apparent hydrodynamic radius $r_{H}$, to a hydrodynamic volume $V_{h}$, and (taking into account the surfactant density and molecular weight) finally to a nominal aggregation number. This procedure was criticized by Kratohvil\cite{kratohvil1980a}, who noted that the hydrodynamic volume of the micelle might well include solvent molecules rather than being composed of pure surfactant. Probe diffusion experiments prove that Kratovil was correct.  The diffusion of probe particles through micellar solutions is retarded by hydrodynamic and direct interactions between the micelles and the probe particles. The degree of retardation is determined by the volume fraction of micelles in the solution.  By combining quasielastic scattering measurements on surfactant solutions and on surfactant-probe mixtures, quasielastic scattering has been used to determine the size, volume fraction, and thus number density of micelles in solution, leading to determinations of the micellar aggregation number and, independently, the (substantial) degree of hydration of micelles, as seen in studies by Phillies and collaborators\cite{phillies1993a}.

(ii) Diffusion of mesoscopic probe particles in  polymer solutions is not Stokes-Einsteinian. $D$ is not determined by the macroscopic viscosity $\eta$.  Therefore, one cannot use the Stokes-Einstein equation for sphere diffusion
\begin{equation}
     D = \frac{k_{B}T}{6 \pi \eta R}
     \label{eq:SEeq}
\end{equation}
(where $\eta$ is the solution viscosity, and $R$ is the sphere radius) to determine the size of probe particles in polymer solutions. However, by using probes of known size, Ullmann and Phillies\cite{ullmann1983a} were able to quantitate the degree of failure of the Stokes-Einstein equation for their polymer solutions, allowing them to measure the size of unknown probe particles in the same solutions.  This approach permitted a quantitative study of the extent of polymer adsorption onto particles chosen for their ability to bind polymers in solution.

\section{Appendix: Isolation of Probe Scattering from Probe-Matrix Systems}

This Appendix discusses experimental approaches used to ensure that spectra of probe-matrix mixtures are correctly interpreted.

First, there is no substitute for checking that the system is not creating experimental artifacts. Spectra of polystyrene spheres in pure water, or monodisperse pure proteins in the presence of added salt (say, 0.15 M) should show close-to-single-exponential relaxations (i.e., nearly straight lines on semilog plots).  The signal-to-noise ratio, the ratio of amplitude at delay time zero to the root-mean-square scatter of the measured spectrum to an adequate fit, should be some number in the range 300 to 1000, or larger.  The data analysis software, no matter what software is used, should confirm that spectra of polystyrene spheres are very close to single exponentials.  A finding that a polystyrene sphere spectrum instead has substantial contributions from relaxation rates differing by 50\% or a factor of 2 indicates that something is wrong with the sphere preparation, the experimental equipment, or the data analysis procedure. There is no point in advancing further until matters are corrected.

Second, there can always be a concern that quasi-stationary features in a complex fluid act effectively as local oscillators, so that detection is actually being made in heterodyne rather than the expected homodyne mode. The effective test is to introduce a local oscillator into the system, so that one has indubitably transitioned from homodyne to heterodyne detection.  An effective scheme for doing this is to lower ever so slowly a screw-mounted needle into the scattering volume, while carefully monitoring the scattering intensity to avoid overloading the detector, until scattering by the needle is much greater than scattering by the solution.  If spectra were original obtained in homodyne mode, then when heterodyne conditions are obtained, the new $S(q,t)$ will have the same time dependence as the $g^{(1)}(q,t)$ obtained before the local oscillator was inserted.

Third, a rapid test checking if matrix scattering is significant is made by using several substantially different (e.g., three-fold different) probe concentrations, to see if spectral fitting parameters are independent of probe concentration\cite{streletzky1998a}.  Some caution is needed. Excessive probe concentrations can lead to multiple scattering, which changes the time dependence of the spectrum.  If spectral fitting parameters are sensitive to matrix concentration, recourse to subtraction at the field correlation level is indicated.

Finally, matrix scattering sometimes contaminates spectra of probes in probe-matrix systems.  This issue is particularly difficult if the matrix spectrum relaxes in whole or part part at shorter times than the probe spectrum relaxes.  In some systems, matrix scattering can be suppressed via spectral subtraction at the field correlation function level. How is subtraction at the field correlation level accomplished?

First, the actual correlation function is generated by breaking time into adjoining intervals of width $\tau$, counting the number of photons received in each interval, and taking crossproducts, namely in the simplest case
\begin{equation}
     S(q, j \tau)   =  \sum_{i=1}^{N} n_{i} n_{i-j}
     \label{eq:correlation function}
\end{equation}

Here $n_{i}$ and $n_{i-j}$  are the numbers of photons counted during time intervals labeled $i$ and $i-j$, the sum goes over the number $N$ of cycle times $\tau$ for which the correlator is operated,  $j$ labels the correlator channels, and for correlators in which all channels have the same width $\tau$ the delay time for channel $j$ is $j \tau$.  For multitau correlators, the relationship between channel widths and delay times is complex\cite{philliesRSI}. The actual $ S(q, j \tau)$ is therefore a large integer.  Some commercial correlators report spectra after applying normalizations to $S(q, j\tau)$. If normalizations are present, they must removed.

Second, confirm that measurements of $g^{(1)}(q,t)$ are stable and reproducible. At large $j$, $ S(q, j \tau)$ decays to a baseline $B$.  There are several paths to determining $B$; they should agree.  $g^{(1)}(q,t)$ is then determined as
\begin{equation}
          g^{(1)}(q,t) =  (S(q, j \tau) - B)^{(1/2)}.
         \label{eq:g1s}
\end{equation}
$t = j \tau$ is the delay time. A first test of system stability is to repeat the measurement several times on the same real sample without changing the sample, laser power, iris and pinhole settings, or photomultiplier tube voltage; the $g^{(1)}(q,t)$ determined repeatedly should agree in amplitude to within experimental error. Also, fitting parameters determined by passing multiple spectra through the data analysis procedure should agree with each other.

Having confirmed that the experimental apparatus is stable, one measures $S(q,t)$ of matrix solution/probe mixtures and also of probe-free matrix solutions.  For a given matrix solution/probe combination, all experimental conditions must be the same in the presence and the absence of probes.  It will usually be convenient to measure the spectrum of the probe-containing mixture first, to ensure that photocount rates due to the probes are acceptable.  Spectra of probe-matrix mixtures and of probe-free matrix solutions are measured, reduced to the corresponding field correlation functions, and subtracted, yielding $g^{(1)}(q,t)$ of the probes.  The same measurements are made several times, on the same system, in order to confirm that the subtracted probe spectra are reproducible.

At this point, several different experimental outcomes are encountered:

\begin{enumerate}

 \item Sometimes refractive index matching of the matrix eliminates matrix scattering. Polyvinylmethylether is index-matched by toluene\cite{cotts1983a} or orthofluorotoluene\cite{lodge1983a}, as demonstrated and vigorously exploited by Lodge and collaborators\cite{lodge1989a}. This approach requires that the (perhaps mixed) solvent simultaneously index-matches the matrix species, is compatible with the matrix species and with the probes, and corresponds to matrix and probe species that are highly monodisperse and available in a wide range of sizes.

\item Sometimes matrix scattering is present but too weak to be significant.  Streletzky and collaborators observed polystyrene spheres in hydroxypropylcellulose: water\cite{streletzky1998a}. Matrix scattering was much weaker than scattering by the spheres.  Even with the smallest sphere studied, at the largest matrix concentrations examined, subtraction of the matrix $g^{(1)}(q,t)$ from the matrix/probe $g^{(1)}(q,t)$ had no effect to within experimental error on spectral fitting parameters. It was therefore not necessary to measure separately and subtract the matrix $g^{(1)}(q,t)$ for every single probe: matrix combination; a sampling showed that this correction was unimportant for the observed probes and matrix concentrations.  In these systems, high-quality spectra of probe-free polymer solutions were obtained by increasing the laser power and the integration time.

\item Sometimes subtraction is mandatory. For polystyrene spheres in solutions of Triton X-100 and other surfactants, light scattering by micelles was significant relative to probe scattering.  Spectral subtraction of surfactant:water spectra was absolutely essential in order to obtain meaningful probe diffusion coefficients\cite{phillies1993a}.

\item Sometimes subtraction is impossible, because matrix scattering is too strong.

\end{enumerate}

\begin{acknowledgments}

I must thank my referees for their suggestions.

\end{acknowledgments}

\end{document}